\documentclass[prd,showpacs,amsmath,amssymb]{revtex4-2}
\usepackage{amsfonts}
\usepackage{bm}
\usepackage{verbatim}
\usepackage{graphicx}
\usepackage{color}
\newcommand{\cor}[1]{{#1}}
\graphicspath{{pics/}}

\begin{document}

\title{Chiral Separation Effect in lattice regularization}

\author{Z.V.Khaidukov}
\affiliation{Institute for Theoretical and Experimental Physics, B. Cheremushkinskaya 25, Moscow, 117259, Russia}

\author{M.A. Zubkov}
\email{zubkov@itep.ru}
\affiliation{Institute for Theoretical and Experimental Physics, B. Cheremushkinskaya 25, Moscow, 117259, Russia}


\date{\today}

\begin{abstract}
We consider Chiral Separation Effect (CSE) in the lattice regularized quantum field theory. We discuss two types of regularization - with and without exact chiral symmetry. In the latter case this effect is described by its conventional expression for the massless fermions. This is illustrated by the two particular cases - of Wilson fermions and of the conventional overlap fermions. At the same time in the presence of the exact chiral symmetry the CSE disappears. This is illustrated by the naive lattice fermions, when the contributions of the fermion doublers cancel each other. Another example is the modified version of the overlap regularization proposed recently, where there is the exact chiral symmetry, but as a price for this the fermion doublers become zeros of the Green function. In this case the contribution to the CSE of zeros and poles of the Green function cancel each other.
\end{abstract}

\maketitle

\section{Introduction}

The family of the non - dissipative transport effects has been widely discussed recently both in the context of the high energy physics and in the context of condensed matter theory \cite{Landsteiner:2012kd,semimetal_effects7,Gorbar:2015wya,Miransky:2015ava,Valgushev:2015pjn,Buividovich:2015ara,Buividovich:2014dha,Buividovich:2013hza}. The possible appearance of such effects in the recently discovered Dirac and Weyl semimetals has been considered \cite{semimetal_effects6,semimetal_effects10,semimetal_effects11,semimetal_effects12,semimetal_effects13,Zyuzin:2012tv,tewary}.
The chiral separation effect (CSE) \cite{Metl} is one of the members of this family. It manifests itself in the equilibrium theory with massless Dirac fermion, where in the presence of external magnetic field (corresponding to the field strength $F_{ij}$) and the ordinary chemical potential $\mu$ the axial current  appears given by
\begin{equation}
j_5^k = -\frac{1}{4\pi^2}\epsilon^{ijk0} \mu F_{ij}\label{1}
\end{equation}
 In the context of the high energy physics the possibility to observe CSE was discussed, in particular, in relation to the  relativistic heavy - ion collisions (see, for example, \cite{Kharzeev:2015znc,Kharzeev:2009mf,Kharzeev:2013ffa} and references therein).

The CSE at a first look has the same origin (the chiral anomaly) as the so - called Chiral Magnetic Effect, which was discussed, for example, in \cite{Vilenkin,CME,Kharzeev:2013ffa,Kharzeev:2009pj,SonYamamoto2012}.
Although it was reported that the possible existence of the  chiral magnetic contribution to ordinary conductivity  \cite{Nielsen:1983rb} was observed in recently discovered Dirac semimetals \cite{ZrTe5}, it was shown that the original equilibrium \cite{Vilenkin} version of CME does not exist. In particular, in \cite{Valgushev:2015pjn,Buividovich:2015ara,Buividovich:2014dha,Buividovich:2013hza} using various numerical methods the CME current was investigated in the context of lattice field theory. It was argued, that the equilibrium bulk CME does not exist, but close to the boundary of the system the nonzero CME current may appear. It was demonstrated, that in the given systems the integrated total CME current remains zero.  In the context of condensed matter theory the absence of CME was reported within the particular model of Weyl semimetal \cite{nogo}. Besides, it was argued, that the equilibrium CME may contradict to the no - go Bloch theorem \cite{nogo2}. However, the way the author of \cite{nogo2} tried to extend the Bloch theorem to the field theoretic systems seems to us non - rigorous. The sufficient analytical proof of the absence of the equilibrium CME (in the systems without superconductivity) was presented by one of us in \cite{Z2016_1,Z2016_2}. This proof relied on the technique of Wigner transformation \cite{Wigner,star,Weyl,berezin}  applied to the lattice regularized quantum field theory.

In the present paper we proceed this line of research and investigate the equilibrium CSE on the same grounds. In the framework of the naive nonregularized quantum field theory the CSE was discussed recently, for example, in \cite{Gorbar:2015wya}. Also it was discussed in the framework of lattice regularization in \cite{Buividovich:2013hza}, where it was argued, that the CSE needs no ultraviolet regularization because the expression for the current does not contain ultraviolet divergences. In the present paper, we argue, however, that the ultraviolet regularization is important. We demonstrate that without it there is an ambiguity in the calculation of the CSE current. Namely, if the model is considered at small but finite temperatures, then the calculation of axial current gives the conventional result if the summation over the Matsubara frequencies is performed first, while if the integration over the $3$ - momenta is performed first, the expression for the current remains undefined. We consider this as an indication, that the rigorous lattice regularization is to be used.
Following the formalism developed in \cite{Z2016_1} in the present paper we consider CSE on the basis of Wigner transformation technique \cite{Wigner,star} applied to the Green functions. Unlike the case of the CME in general case the coefficient in the linear response of the axial current to the external magnetic field is not a topological invariant. However, it appears, that the coefficient in the CSE current standing at the product of magnetic field and chemical potential approaches  the topological invariant when mass of the fermion tends to zero. This allows us to derive the conventional expression for the CSE current.  Thus the link between the CSE effect in lattice regularization and momentum space topology is established.

It is worth mentioning, that Momentum space topology is a powerful method, which was developed earlier mainly within condensed matter theory. It allows to describe in a simple way, for example, the stability of the Fermi points, the Anomalous Quantum Hall effect, and the fermion zero modes on vortexes (for the review see \cite{Volovik2003,Volovik:2011kg}). Recently certain aspects of momentum space topology were discussed in the framework of the four - dimensional lattice gauge theory (see for example \cite{VZ2012,Z2012,Z2016_3}).

The paper is organized as follows. In Sect. \ref{SectWigner} following \cite{Z2016_1,Z2016_2} we describe how the technique of the Wigner transformation may be applied to the lattice regularized quantum field theory. In Sect. \ref{SectLat} we consider the CSE in conventional lattice regularizations (Wilson fermions and overlap fermions). In Sect. \ref{SectLat2} we consider the CSE in the case of exact chiral symmetry - for the naive lattice regularization with $16$ doublers and in the lattice regularization with deformed overlap fermions, where instead of the $15$ doublers the zeros of the Green function appear. In Sect. \ref{SectCont} we demonstrate the ambiguity in the calculation of the CSE current that takes place in the naive continuum theory. In Sect. \ref{SectConcl} we end with the conclusions.

\section{Lattice fermions in the presence of external $U(1)$ gauge field}
\label{SectWigner}
\subsection{Lattice models in momentum space}

In this section we consider briefly the lattice models in momentum space following the methodology of \cite{Z2016_1,Z2016_2}. For the more detailed description of the method and for the references see \cite{Z2016_1,Z2016_2}.
In the absence of the external gauge field  the partition function of the theory defined on the infinite lattice may be written as
\begin{equation}
Z = \int D\bar{\psi}D\psi \, {\rm exp}\Big( - \int_{\cal M} \frac{d^D {p}}{|{\cal M}|} \bar{\psi}^T({p}){\cal G}^{-1}({ p})\psi({p}) \Big)\label{Z1}
\end{equation}
where $|{\cal M}|$ is the volume of momentum space $\cal M$, $D$ is the dimensionality of space - time, while $\bar{\psi}$ and $\psi$ are the Grassmann - valued fields defined in momentum space $\cal M$. $\cal G$ is specific for the given system. For example, the model with $3+1$ D Wilson fermions corresponds to $\cal G$ that has the form
 \begin{equation}
 {\cal G}({p}) = - i \Big(\sum_{k}\gamma^{k} g_{k}({p}) - i m({p})\Big)^{-1}\label{G10}
 \end{equation}
 where $\gamma^k$ are Euclidean Dirac matrices
 defined in chiral representation
$$
\gamma^4 = \left(\begin{array}{cc} 0 & 1 \\1&0 \end{array}\right), \quad \gamma^i = \left(\begin{array}{cc} 0 & i\sigma^i\\- i\sigma^i&0 \end{array}\right), \quad i = 1,2,3$$ $$\gamma^5 = \left(\begin{array}{cc} 1 & 0\\0&-1 \end{array}\right)
$$
where $\sigma^i$ is the Pauli matrix.
$g_k({p})$ and $m({p})$ are the real - valued functions ($k = 1,2,3,4$) given by
\begin{equation}
g_k({p}) = {\rm sin}\,
p_k, \quad m({p}) = m^{(0)} +
\sum_{a=1,2,3,4} (1 - {\rm cos}\, p_a)\label{gWilson}
\end{equation}
The fields in coordinate space are related to the fields in momentum space as follows
\begin{equation}
\psi({r}) = \int_{\cal M} \frac{d^D {p}}{|{\cal M}|} e^{i {p}{r}} \psi({p})\label{Psip}
\end{equation}
At the discrete values of $r$ corresponding to the points of the lattice this expression gives the values of the fermionic field at these points, i.e. the dynamical variables of the original lattice model. However, Eq. (\ref{Psip}) allows to define formally the values of fields at any other values of $r$. The partition function may be rewritten in the form
\begin{equation}
Z = \int D\bar{\psi}D\psi \, {\rm exp}\Big( - \sum_{{r}_n} \bar{\psi}^T({r}_n)\Big[{\cal G}^{-1}(-i\partial_{r})\psi({ r})\Big]_{{r}={r}_n} \Big)\label{Z2}
\end{equation}
Here the sum in the exponent is over the discrete coordinates ${r}_n$. However, the operator $-i\partial_{r}$ acts on the function $\psi({r})$ defined using Eq. (\ref{Psip}). In order to derive Eq. (\ref{Z2}) we use identity
\begin{equation}
\sum_{r}e^{i{p}{r}} = |{\cal M}|\delta({p})
\end{equation}
In the particular case of Wilson fermions we may rewrite the partition function in the conventional way as
\begin{equation}
Z = \int D\bar{\psi}D\psi \, {\rm exp}\Big( - \sum_{{r}_n,{r}_m} \bar{\psi}^T({r}_m)({\cal D}_{{r}_n,{r}_m}) \psi({r}_n) \Big)\label{Z28}
\end{equation}
with
\begin{equation}
{\cal D}_{{x},{y}}  =  - \frac{1}{2}\sum_i [(1 +
\gamma^i)\delta_{{x}+{e}_i, {y}}  +  (1 -
\gamma^i)\delta_{{x}-{e}_i, {y}} ] +  (m^{(0)} + 4)
\delta_{{x}{y}}\label{Wilson}
\end{equation}
Here ${e}_i$ is the unity vector in the $i$ - th direction.

\subsection{Introduction of the gauge field}

Gauge transformation of the lattice field takes the form
\begin{equation}
\psi({r}_n)\rightarrow e^{i \alpha({r}_n)} \psi({r}_n)\label{gt}
\end{equation}
In case of Wilson fermions the $U(1)$ gauge field is typically introduced as the following modification of operator $D$:
\begin{eqnarray}
{\cal D}_{{x},{y}} & = & - \frac{1}{2}\sum_i [(1 +
\gamma^i)\delta_{{x}+{e}_i, {y}}e^{i A_{{x}+{e}_i, {y}}} \nonumber\\&& +  (1 -
\gamma^i)\delta_{{x}-{e}_i, {y}} e^{i A_{{x}-{e}_i, {y}}}] +  (m^{(0)} + 4)
\delta_{{x}{y}}\label{WF}
\end{eqnarray}
Here $A_{{x}, {y}} = - A_{{y}, {x}}$ is the gauge field attached to  the links of the lattice. In the same way the gauge field is typically incorporated into the models of solid state physics.

One can easily check, that Eq. (\ref{Z28}) may be rewritten as
\begin{eqnarray}
Z = \int D\bar{\psi}D\psi \, {\rm exp}\Big( -  \int_{\cal M} \frac{d^D {p}}{|{\cal M}|} \bar{\psi}^T({p})\hat{\cal Q}(i{\partial}_{p},{p})\psi({p}) \Big)\label{Z4}
\end{eqnarray}
Here
\begin{equation}
\hat{\cal Q} = {\cal G}^{-1}({p} - {A}(i{\partial}^{}_{p}))\label{calQM}
\end{equation}
while the pseudo - differential operator ${A}(i\partial_{p})$ is defined as follows. First, we represent the original gauge field ${A}({r})$ as a series in powers of coordinates ${r}$. Next, variable ${r}$ is substituted in this expansion by the operator $i\partial_{p}$. Besides, in Eq. (\ref{calQM}) each product of the components of ${p} - {A}(i{\partial}^{}_{p})$ is subsitituted by the symmetric combination (for the details see \cite{Z2016_1}). As it was mentioned above, for the case of Wilson fermions the formulations of Eq. (\ref{Z4}) and Eq. (\ref{Z28}) are exactly equivalent. For the other regularizations there may be the difference, but it manifests itself in the terms that are proportional to the field strength times $a^2$ (here $a$ is the lattice spacing). Those extra terms may be neglected in continuum limit. Therefore, for {\it any} regularization we accept Eq. (\ref{calQM}) as the definition of the model in the presence of external $U(1)$ gauge field.

\subsection{Electric current}

Electric current is defined as the response of effective action $-{\rm log}\, Z$ to the variation of external Electromagnetic field. This gives \cite{Z2016_1}
\begin{eqnarray}
j^k({R}) &=& \cor{-}\int_{\cal M} \frac{d^D {p}}{(2\pi)^D} \,  {\rm Tr} \, \tilde{G}({R},{p}) \frac{\partial}{\partial p_k}\Big[\tilde{G}^{(0)}({R},{p})\Big]^{-1}\label{j423}
\end{eqnarray}
where the Wigner transformation of the Green function is expressed as:
\begin{equation}
 \tilde{G}({R},{p}) = \sum_{{r}={r}_n} e^{-i {p} {r}} G({R}+{r}/2,{R}-{r}/2)\label{Wl2}
\end{equation}
while the Green function itself is
\begin{eqnarray}
G({r}_1,{r}_2)&=& \cor{-}\frac{1}{Z}\int D\bar{\Psi}D\Psi \,\bar{\Psi}({r}_2)\Psi({r}_1)\\ &&{\rm exp}\Big(-\sum_{{ r}_n}\Big[ \bar{\Psi}({r}_n)\Big[{\cal G}^{-1}(-i\partial_{r}  \nonumber\\&& - {A}({r}))\Psi({r})\Big]_{{r}={r}_n}\Big]\Big)\nonumber
\end{eqnarray}
At the same time
\begin{eqnarray}
&&\tilde G^{(0)}({R},{p})  = {\cal G}({p}-{A}({R}))\label{Q0}
\end{eqnarray}
The method of Wigner transformation in the application to the lattice models was developed in \cite{Z2016_1,Z2016_2} following its original formulation specific for the theory in continuous space - time \cite{Wigner,star,Weyl,berezin}.
In \cite{Z2016_1} the following expression was derived for the linear response of the electric current to external electromagnetic field: \begin{eqnarray}
j^{(1)k}({R})  &= &\cor{-} \frac{1}{4\pi^2}\epsilon^{ijkl} {\cal M}_{l} A_{ij} ({R}), \label{calM}\\
{\cal M}_l &=& \int_{} \,{\rm Tr}\, \nu_{l} \,d^4p \label{Ml} \\ \nu_{l} & = &  - \frac{i}{3!\,8\pi^2}\,\epsilon_{ijkl}\, \Big[  {\cal G} \frac{\partial {\cal G}^{-1}}{\partial p_i} \frac{\partial  {\cal G}}{\partial p_j} \frac{\partial  {\cal G}^{-1}}{\partial p_k} \Big]  \label{nuG}
\end{eqnarray}

\section{Lattice regularization with broken chiral symmetry}
\label{SectLat}
\subsection{Linear response of chiral current to external magnetic field}

In this section we consider the linear response of the chiral current to external Electromagnetic field. For the field system in continuous coordinate space this response may easily be calculated using Feynman diagrams. For the field system in lattice regularization this response may be calculated following the approach of \cite{Z2016_1,Z2016_2,Z2016_3} that was briefly described above.
In continuum theory the naive expression for the chiral current is $\langle \bar{\psi} \gamma^\mu \gamma^5 \psi\rangle$. Several different definitions for the particular lattice regularization may give this expression in the naive continuum limit.

The evident choice of the definition of current in lattice regularization is the functional derivative over axial gauge field of the effective action. The latter field, in turn, may be defined through the covariant derivative, which acts on the left - handed and the right - handed fermions via opposite charges. For the particular choice of the lattice model with exact chiral symmetry (i.e. when $\cal G$ commutes or anti - commutes with $\gamma^5$) this definition gives the expression similar to that of \cite{Z2016_1,Z2016_2,Z2016_3}:
\begin{eqnarray}
j^{5k}({R}) &=& \int_{\cal M} \frac{d^D {p}}{(2\pi)^D} \,  {\rm Tr} \,\gamma^5\,  \tilde{G}({R},{p}) \frac{\partial}{\partial p_k}\Big[\tilde{G}^{(0)}({R},{p})\Big]^{-1}\label{j423}
\end{eqnarray}
where
\begin{eqnarray}
&&\tilde G^{(0)}({R},{p})  = {\cal G}({p}-\cor{A}({R}))\label{Q01}
\end{eqnarray}
Actually, we are able to adopt this definition to any lattice theory even without the exact chiral symmetry. One can easily check that in the naive continuum limit this definition gives $\langle \bar{\psi} \gamma^k \gamma^5 \psi\rangle$.

In order to regularize our expressions for the case of truly massless fermions let us use the finite temperature version of the lattice theory. With the periodic boundary conditions in the spatial directions
and anti-periodic in the imaginary time direction, the lattice momenta will be
\begin{equation}\label{disc}
p_i \in (0, 2\pi);\, p_4=\frac{2\pi}{N_t }(n_4+1/2)
\end{equation}
where $i=1,2,3$ while $n_4=0,...,N_t-1$. Temperature is equal to $T = 1/N_t $, in lattice units $1/a$, where $a$ is the lattice spacing. Thus the imaginary frequencies are discrete $p_4 = \omega_{n}=2\pi T (n+1/2)$, where $n= 0, 1, ... N_t-1$, while the axial current (also in lattice units) is expressed via the Green's functions as follows:
\begin{eqnarray}
j^{5k}&=&-\frac{i}{2}T\sum_{n=0}^{N_t-1}\int \frac{d^3p}{(2\pi)^3}{\rm Tr}\, \gamma^5 ({\cal G}(\omega_{n},\textbf{p})\partial_{p_{i}}{\cal G}^{-1}(\omega_{n},\textbf{p})\nonumber \\&&\partial_{p_{j}}{\cal G}(\omega_{n},\textbf{p})\partial_{p_{k}}{\cal G}^{-1}(\omega_{n},\textbf{p}))F_{ij}\label{24}
\end{eqnarray}

\subsection{The linear response of the chiral current to chemical potential and to external magnetic field}

Now let us consider the system without exact chiral symmetry. Recall, that the exact chiral symmetry is to be broken if we want to describe one Dirac fermion, which is related to Nielsen - Ninomiya theorem. For definiteness, we may discuss first Wilson fermions. But this is not necessary, and the results of this subsection are valid for any lattice models. We introduce the chemical potential in the standard way $\omega_{n} \to \omega_{n}-i\mu$. The derivative of the current with respect to $\mu$ gives
\begin{equation}
j^{5k}= \frac{{\cal N}^{ijk}}{4\pi^2} F_{ij} \mu\label{jmuH}
\end{equation}
with
\begin{eqnarray}
{\cal N}^{ijk}&=&-\sum_{n=0}^{N_t-1}\frac{1}{2 }T\int \frac{d^3p}{(2\pi)}\partial_{\omega_n}{\rm Tr}\, \gamma^5 {\cal G}(\omega_{n},\textbf{p})\partial_{p_{i}}{\cal G}^{-1}(\omega_{n},\textbf{p})\nonumber \\&&\partial_{p_{j}}{\cal G}(\omega_{n},\textbf{p})\partial_{p_{k}}{\cal G}^{-1}(\omega_{n},\textbf{p})\label{26}
\end{eqnarray}

We assume that the singularities of the Green function may appear only at the finite sequence of values of $\omega = \omega^{(0)}, \omega^{(1)}, ... $ that do not coincide with the Matsubara frequencies. In the limit $T \to 0$ the sum over Matsubara frequencies becomes the integral that is regularized as follows
\begin{equation}
{\cal N}^{ijk}= \sum_k \Big(- {\cal N}^{ijk}_3(\omega^{(k)}+0) + {\cal N}^{ijk}_3(\omega^{(k)}-0)\Big)
\end{equation}
where ${\cal N}_3(\omega_n)$ is given by
\begin{eqnarray}
{\cal N}^{ijk}_{3}(\omega_n)&=&-\frac{1}{2} \int \frac{d^3p}{(2\pi)^2}{\rm Tr}\, \gamma^5 {\cal G}(\omega_{n},\textbf{p})\partial_{p_{i}}{\cal G}^{-1}(\omega_{n},\textbf{p}) \nonumber \\&& \partial_{p_{j}}{\cal G}(\omega_{n},\textbf{p}) \partial_{p_{k}}{\cal G}^{-1}(\omega_{n},\textbf{p})
\end{eqnarray}
It is clear, that in the model, in which there are no poles or zeros of the Green function in the presence of exact chiral symmetry the linear response of the axial current to magnetic field is the sum of topological invariants, i.e. it cannot be changed under the continuous deformations of the model. However, in general case when $\gamma^5$ does not (anti) commute with the Green function, the terms in this expansion are not topological invariants.
We may also rewrite
\begin{eqnarray}
{\cal N}^{ijk}&=&\frac{1}{2}\int_{\Sigma}\frac{d^3p}{(2\pi)^2}{\rm Tr}\, \gamma^5 {\cal G}(\omega_{},\textbf{p})\partial_{p_{i}}{\cal G}^{-1}(\omega_{},\textbf{p})\nonumber \\&&\partial_{p_{j}}{\cal G}(\omega_{},\textbf{p}) \partial_{p_{k}}{\cal G}^{-1}(\omega_{},\textbf{p})\label{calN}
\end{eqnarray}
where $\Sigma$ is the 3D hypersurface of infinitely small volume that embraces the singularities of the Green function concentrated at the Fermi surfaces (or Fermi points). \cor{It has the form of the two infinitely close pieces situated at $\omega = \pm 0$.} The advantage of this representation is that Eq. (\ref{calN}) becomes the topological invariant if $\gamma^5$ anti - commutes with the Green function in a small vicinity of its poles. \cor{Namely, in this case we may represent
\begin{eqnarray}
{\cal N}^{[ij]k}&=&\frac{\epsilon^{ijk}}{ 3!}\int_{\Sigma}\frac{1}{(2\pi)^2}{\rm Tr}\, \gamma^5 {\cal G}(\omega_{},\textbf{p})d {\cal G}^{-1}(\omega_{},\textbf{p})\nonumber \\&&\wedge d{\cal G}(\omega_{},\textbf{p})\wedge d{\cal G}^{-1}(\omega_{},\textbf{p})\nonumber
\end{eqnarray}
Here $\Sigma$ is a small closed surface in $4D$ space. It may have an arbitrary form surrounding the singularities.}

\subsection{Regularization with Wilson fermions}

For the case of Wilson fermions the singularities of the Green function may appear at $\omega = 0, \pi$ (for $m^{(0)}>0$ they appear at $\omega = 0 $ only). Then the limit $T\to 0$ gives
\begin{equation}
{\cal N}^{ijk}= - {\cal N}^{ijk}_3(+0) + {\cal N}^{ijk}_3(-0) + {\cal N}^{ijk}_3(\pi-0) - {\cal N}^{ijk}_3(\pi+0)
\end{equation}

The interesting particular case is when parameter $m^{(0)}$ vanishes. In this case at $\mu = 0$ the only Fermi point appears at ${\bf p}=0$ and on $\Sigma$ we have $\{\gamma^5,{\cal G}\}\approx 0$. In this particular case $${\cal N}^{ijk} = \epsilon^{ijk}$$ which gives the regular expression for the Chiral Separation Effect of Eq. (\ref{1}). In order to confirm this prediction we use also the numerical methods. Namely, we take Eq. (\ref{26}) and calculate numerically for the component ${\cal N}^{123}$ the integral over $3$ - momenta and the sum over $\omega_n$ using MAPLE package. It is seen, that at $N_t \to \infty$ the answer tends to $1$ as it should.

In the presence of nonzero mass $m^{(0)}>0$ the situation is changed, and the poles of the Green function do not appear while $\mu < m^{(0)}$, which gives the vanishing CSE current. At $\mu \ge m^{(0)}$ the Fermi surface appears, and it contributes to the chiral current through Eq. (\ref{calN}). However, in this case $\gamma^5$ does not anti commute with $\cal G$ on $\Sigma$. In the continuum limit $m^{(0)} = m^{(0)}_{\rm phys} a$ and $\mu = \mu_{\rm phys} a$, where $m^{(0)}_{\rm phys} $ and $\mu_{\rm phys}$ are the parameters of the model in physical units while $a$ is the lattice spacing. In continuum limit $a \to 0$ at $\mu_{\rm phys}  \gg m^{(0)}_{\rm phys}$ we recover the conventional result for the CSE of Eq. (\ref{1}).

\subsection{Overlap fermions}

Let us discuss the regularization using overlap fermions \cite{overlap}. The massless overlap Dirac operator is defined as
\begin{equation}
{\cal D}_o=m(\hat{1}+ {\cal D}(-m)\,({\cal D}(-m){\cal D}^+(-m))^{-1/2})
  \end{equation}
where ${\cal D}(m^{(0)})$ is the dimensionless  Wilson-Dirac operator given by Eq. (\ref{Wilson}), where we substitute negative value of the mass parameter $m^{(0)} = - m$.
Operator ${\cal D}_o$  obeys Ginsparg - Wilson relation, which may be written in the following form
\begin{equation}
\{{\cal D}^{-1}_o,\gamma^5\} = \frac{\gamma^5}{m}
\end{equation}
It is called sometimes $"$ exact $"$   chiral symmetry on the
lattice. However, this statement is not precise, and actually the conventional overlap propagator does not obey the exact chiral symmetry, which is $\{{\cal D}^{-1}_o,\gamma^5\} =0$.

In momentum space we have
\begin{equation}
{\cal D}^{-1}_o=-i\gamma_{\mu}C_{\mu}+\frac{1}{2 m},
\end{equation}
with
\begin{equation}C_{\mu}(p)=\frac{1}{2m}\frac{k_{\mu}}{\sqrt{k_{\mu}^2+A^2}+A},A=\frac{\hat{k}_{\mu}^2}{2}- m
\end{equation}
and
\begin{equation} k_{\mu}={\rm sin}(p_{\mu}),\hat{k}=2{\rm sin}(\frac{p_\mu}{2}) \end{equation}

In this model the fermion Green function $${\cal G} = {\cal D}_o^{-1}$$
has the only pole at $\omega = 0, {\bf p}=0$. At $p = (n_1\pi, n_2\pi, n_3 \pi, n_4 \pi)$ with integer $n_i = 0,1$ such that $n_1+n_2+n_3+n_4\ne 0$ we have the value ${\cal G}(p) = \frac{1}{2m}$.

Again, the above consideration may be applied to the model in this regularization, and we have the expression for the linear response to the magnetic field given by Eq. (\ref{jmuH}) with $\cal N$ of Eq. (\ref{calN}). In particular, in the continuum limit when $m \to 0$, and $\mu = 0$ we get ${\cal N}^{ijk} = \epsilon^{ijk}$ that results in the usual expression for the CSE current of Eq. (\ref{1}).

\section{Lattice regularization with exact chiral symmetry}
\label{SectLat2}
\subsection{Naive lattice fermions}

In this section we consider the case of the lattice model with exact chiral symmetry, when
$$\{\gamma^5,{\cal G}\}=0$$
The simplest example of such a system is given by the naive lattice fermions with the Green function in momentum space of the form
 \begin{equation}
 {\cal G}({p}) = - i \Big(\sum_{k}\gamma^{k} g_{k}({p}) - i m^{(0)}\Big)^{-1}\label{G102}
 \end{equation}
 where $\gamma^k$ are Euclidean Dirac matrices while $g_k({p})$  are the real - valued functions ($k = 1,2,3,4$) given by
\begin{equation}
g_k({p}) = {\rm sin}\,
p_k \label{gnaive}
\end{equation}
In this model at $m^{(0)} = 0$ instead of one massless Dirac particle in continuum limit there are $16$ massless particles. In this case the linear response of the chiral current to external magnetic field and chemical potential is given by
Eq. (\ref{jmuH}) with $\cal N$ of Eq. (\ref{calN}).
The contributions of the doublers differ due to the orientation of effective vierbein, i.e. the corresponding low energy effective theory for massless particle has the one - particle Euclidean lagrangian
$$
{\cal L} = |e| \,e^\mu_a \gamma^a i \nabla_\mu
$$
where
$$
|e| \, e^\mu_a = \left(\begin{array}{cccc}(-1)^{n_1} & 0 & 0 & 0 \\
0 &(-1)^{n_2} & 0 & 0\\
0 & 0 & (-1)^{n_3} & 0\\
0 & 0 & 0 & (-1)^{n_4}
 \end{array}\right)
$$
with $n_i = 0,1$. As a result the contributions to CSE current of those $16$ doublers cancel each other.
Thus, unlike the case of the previous section, in the continuum limit when $m \to 0$, and $\mu = 0$ all doublers contribute the sum thus giving ${\cal N} = 0$, and cancelling the overall CSE current.

\subsection{Modified overlap fermions}

It was proposed (see, for example, \cite{overlap}) to redefine the overlap fermions as follows
$${\cal G}={\cal D}_o^{-1}-\frac{1}{2 m}$$
In this case the chiral symmetry is exact
\begin{equation}
\{{\cal G},\gamma^5\}=0\label{exactchiral}
\end{equation}
But the price for this is that
at $p = (n_1\pi, n_2\pi, n_3 \pi, n_4 \pi)$ with $n_1+n_2+n_3+n_4\ne 0$ we have the vanishing value of the Green function ${\cal G}(p) = 0$.

The zeros of the Green function are in many aspects similar to poles. In particular, they contribute to CSE in such a way, that the total current vanishes.
Let us define
$$f(k^2)=\frac{2 m (\sqrt{k^2+A^2}+A)}{k^2}$$
Then
\begin{equation}
{\cal G}=-i\frac{k_{\mu}\gamma^{\mu}}{f(k^2)k^2}  \label{eqs}
\end{equation}

At finite temperatures we have
\begin{equation}
j^{5k}=\frac{-iT}{2\pi}\sum_{n=0}^{N_t-1}{\cal N}_{3}(\omega_n) \epsilon^{ijk} F_{ij}
\end{equation}
At any value of $n$ the functional ${\cal N}_3(\omega_n)$ is the topological invariant, i.e,
it is not changed under any variation ${\cal G} \to {\cal G}+\delta {\cal G}$ if during such modification the poles or zeros of $\cal G$ do not appear. It is given by
\begin{eqnarray}
{\cal N}_{3}(\omega_n)&=&\frac{1}{2\times 3!}\epsilon^{ijk}\int\frac{d^3p}{(2\pi)^2}{\rm Tr} \, \gamma^5 {\cal G}(\omega_{n},\textbf{p})\partial_{p_{i}}{\cal G}^{-1}(\omega_{n},\textbf{p})\nonumber \\&&\partial_{p_{j}}{\cal G}(\omega_{n},\textbf{p})\partial_{p_{k}}{\cal G}^{-1}(\omega_{n},\textbf{p})
\end{eqnarray}
Therefore, in the model, in which there are no poles or zeros of the Green function in the presence of exact chiral symmetry the linear response of the axial current to magnetic field is the sum of topological invariants, i.e. it cannot be changed under the continuous deformations of the model. The pole of the Green function at finite temperature may appear if there exists such integer $n$ that $\omega_n = \frac{2\pi}{N_t}(n + 1/2)  = \pi$. This gives  equation $$2n + 1 = N_t$$ which  has a solution for odd values of $N_t$. Therefore, for simplicity in the following we assume that $N_t$ is even.

The ordinary chemical potential cannot cause the appearance of poles or zeros of the Green function, which is seen from the following consideration. Again, let us assume that the chemical potential appears as the imaginary contribution to Matsubara frequency. In this case the poles or zeros of the Green function may appear if:
\begin{equation}
{\rm sin}^{2}(\omega_n-i\mu)+\sum_{l=1}^3 {\rm sin}^{2}(p_{i})=0
\end{equation}
We obtain the following system
\begin{equation}
\begin{cases}
1-{\rm cos}(2\omega_n) {\rm ch}(2\mu)+2\sum_{l=1}^{3}{\rm sin}^2(p_{i})=0\\
{\rm sh}(\mu)\,{\rm sin}(2\omega_n)=0\\
\end{cases}
\end{equation}
The second equation has a solution in real variables. We obtain $\omega_n= \pi/2 $ or $\omega_n=\pi$. In the former case the first equation does not have solutions. The latter case is realized for odd values of $N_t$ only, and then the poles of the Green function appear as the solution of equation
$$
1+2\sum_{l=1}^{3}{\rm sin}^2(p_{i})={\rm ch}(2\mu)
$$
However, as above we can always choose the even value of $N_t$, and therefore the poles of the Green function do not appear if we modify the value of $\mu$. Therefore, we are able to calculate ${\cal N}_3(\omega_n)$ for vanishing $\mu$, and the result gives the answer for finite $\mu$. This calculation is represented in Appendix A, and as expected it gives the vanishing CSE current.

\section{Naive continuum expressions for the CSE current}
\label{SectCont}
\subsection{The integration over $3$ - momenta before the summation over Matsubara frequencies}

Above we considered the CSE effect using rigorous lattice regularizations. Also we feel this instructive to present here the discussion of chiral separation effect in the framework of naive continuous field theory. We will see, that there is an ambiguity in this consideration, which is reflected by the lattice constructions with and without exact chiral symmetry.

Let us  consider  the propagator of massless non-interacting Dirac fermions
\begin{equation}
G(\omega_{n},\textbf{p})=\frac{1}{\gamma^{\mu}p_{\mu}}
\end{equation}
Here $p_\mu = (\omega,{\bf p})$. We will consider the case when the magnetic field is directed along the z axis (i.e $F_{12}=-B_{}$). The expression for the chiral current has the form:
\begin{equation}\label{summ}
j^{5z}=4Ti\sum_{n=-\infty}^{\infty}\int{\frac{d^{3}p}{(2\pi)^3}\frac{\omega_{n}}{((\omega_{n})^2+p^2)^2}}B_{}
\end{equation}
and after integration over $3$ - momenta we arrive at
\begin{equation} j^{5z}=4Ti\pi^2\sum_{n=-\infty}^{\infty}{\rm sign}({\omega_{n}})B_{}\label{sumo} \end{equation}
Thus, formally, in the case of massless fermions the chiral current is equal to the sum of the integer numbers. If, as a result of the interaction in medium, $G$ is changed as $\omega_{n} \to f(\omega_{n}),p_{i} \to g(p_{i}) $,  then the result depends on ${\rm sign}\,f$.

We introduce the chemical potential in the standard way $\omega_{n} \to \omega_{n}-i\mu$. In this case  we need  an analytical continuation of the sign function. We may try to use, for example, the rule \begin{equation} {\rm sign}(\omega_{n}-i\mu)={\rm sign}({\rm Re}(\omega_{n}-i\mu))\end{equation}
And then from this naive consideration the conclusion may be drawn, that the chemical potential does not influence the CSE current.
Below we will see, that the formal expressions in the continuum theory will lead to the different answer if the summation over the Matsubara frequencies is performed before the integration over momenta.

\subsection{The integration over $3$ - momenta after the summation over Matsubara frequencies}

In the previous section we have shown that the axial current in an external magnetic field is expressed as the sum of the topological invariants multiplied by the field strength. This expression would become the exact result, but only if the theory does not contain divergences.

We may extract another result from the above expressions. Namely, let us first perform the initial summation over the frequencies and only after that - the integration over momenta
\begin{equation}j^{5z}=4Ti\sum_{n=-\infty}^{\infty}\int{\frac{d^3p}{(2\pi)^3}\frac{\omega_{n}-i\mu}{((\omega_{n}-i\mu)^2+p^2)^2}}B=
\end{equation}
\begin{equation}
=2\int_{C}{\frac{dz}{2i\pi}\int{\frac{d^3p}{(2\pi)^3}\frac{z}{(z^2-p^2)^2}{\rm th}(\frac{z-\mu}{T})}B}\label{49}
\end{equation}
Where $C$ is the contour that surrounds poles of the hyperbolic tangent function.
We use the relation
\begin{equation}{\rm res}_{z=z0}f(z)=\frac{1}{(m-1)!}\lim_{z \to z_0}\frac{d^{m-1}}{dz^{m-1}}f(z)(z-z_0)^m
\end{equation}
  to calculate the value of the integral using the theory of residues. We deform contour $C$ in such a way that it surrounds the points $z=\pm z_0$.
	 After this deformation we have
	\begin{eqnarray}
	j^{5z}&=&- 2\int{\frac{d^3p}{(2\pi)^3}}\Big[\frac{z}{(z+p)^2}\frac{d}{dz}{\rm th}(\frac{z-\mu}{2T})|_{z=p}\nonumber \\&&+
	\frac{z}{(z-p)^2}\frac{d}{dz}{\rm th}(\frac{z-\mu}{2T})|_{z=-p}\Big]
		\end{eqnarray}
	We can write the equation in this form because:
	\begin{equation}\frac{d}{dz}\frac{z}{(z \mp p)^2}=0\end{equation}
	Thus we can rewrite the integral as 	\begin{equation}j^{5z}=-2B\int{\frac{d^3p}{(2\pi)^3}}(\frac{1}{4p}\frac{d}{dp}{\rm th}(\frac{p-\mu}{2T})-\frac{1}{4p}\frac{d}{dp}{\rm th}(\frac{p+\mu}{2T}))\end{equation} and  after substitution $d^3p=4\pi p^2dp$ we find that
	\begin{equation}j^{5z}=-\frac{B}{2\pi^2}\int{dp(n_f(p-\mu)-n_f(p+\mu))}=-\frac{B\mu}{2\pi^2}
	\end{equation}
This expression coincides with the conventional expression Eq. (\ref{1}) (see, for example, \cite{Metl}) and also it coincides with the result obtained above using Wilson fermions and conventional overlap fermions. It is still protected from the renormalization of $3$ - momentum ($p_i \to g(p_i)$).

However, although the result of the rigorously regularized theory (using, say,  lattice Wilson fermions) is reproduced by the approach of the present subsection, we would like to emphasize once again, that this approach itself is not self - consistent, and its application to the other problems may be limited. In particular, let us consider the modification of the system, which leads to the replacement of $i\omega_{n}$ by a function $f(i\omega_n)$ such that it tends to  $i\omega_{n}$  at large $n$. Looking at Eq. (\ref{summ}) we may come to the conclusion, that such a modification cannot change the value of the axial current. However, Eq. (\ref{49}) is not invariant under the substitution $i\omega_{n} \to f(i\omega_n)$.
This demonstrates once again, that the rigorous ultraviolet regularization is needed in order to calculate the response of axial current to external field strength.

\section{Conclusions and discussions}
\label{SectConcl}
In the present paper we discuss the Chiral Separation Effect both in the framework of the naive continuum nonregularized quantum field theory and of the lattice regularized theory.
In both cases we also regularize the theory using finite temperatures.

We demonstrate, that the naive continuum formulation suffers from the ambiguities related to the order of taking the integral over the $3$ - momenta and the sum over Matsubara frequencies. If the Matsubara frequencies are summed first, then the divergencies are not encountered and the conventional expression for the chiral current in the presence of external magnetic field is reproduced. At the same time if the $3$ - momenta are integrated first, then the resulting expression is given by the sum of Eq. (\ref{sumo}), where each term is equal to either $1$ or $-1$. This sum is not well defined, but each term in this sum is indepedent of the chemical potential.

This ambiguity points out that although certain computational schemes of the CSE current do not encounter the ultraviolet divergences, the theory should be considered in the ultraviolet regularization in order to obtain rigorous results. Therefore, we consider several types of lattice regularization. First of all, we considered the naive lattice regularization, where $16$ doublers represent the independent physical excitations. Those excitations differ by the orientation of effective vierbein, and as a result their contributions to the CSE current cancel each other.

Recently the modification of the regularization using overlap fermions was proposed (see, for example, \cite{overlap}), in which massless physical excitation appears at $\omega = {\bf p} = 0$ while at the positions of the other $15$ doublers (of naive lattice fermions), the zeros of the Green function appear. As a result the exact lattice chiral symmetry is obeyed just like in the case of naive lattice fermions. The physical meaning of the zeros of the Green function remains unclear, but it is discussed in certain publications (mostly in the framework of condensed matter theory). We demonstrate, that the contribution of those zeros of the Green function to the CSE cancels the contribution of the physical massless excitation.
Thus in both considered cases of the lattice theory with exact chiral symmetry the CSE does not appear.

Typically, in the lattice models the exact chiral symmetry is broken, which is the price for the elimination of the fermion doublers. We consider the two particular cases of such conventional regularization - the case of lattice Wilson fermions and the case of conventional overlap fermions. In both cases the massless excitation appears at $\omega = {\bf p} = 0$ only, the other doublers disappear, and also there are no zeros of the Green function. The price for this is the absence of the exact chiral symmetry. However, in the case of overlap fermions there is the Ginsparg - Wilson relation instead.
In both these regularizations we observe the emergence of the Chiral Separation Effect. The corresponding current tends to its conventional expression Eq. (\ref{1}) in continuum limit of the model with massless fermions. In the case, when the theory describes massive fermions with mass $m$, the CSE current is absent at $\mu < m$. It appears at $\mu \ge m$, and is given by the same expression of Eq. (\ref{1}) in the limit $\mu \gg m$.

Actually, our consideration may easily be extended to the other lattice models including those ones, with the interactions. The necessary condition is the presence of the massless Dirac fermions in continuum limit. Therefore, Eq. (\ref{1}) should be regularization independent. We take the limit $T \to 0$, which allows to substitute the sum over Matsubara frequencies by the integral. This consideration demonstrates also, that for the noninteracting system the same answer for the CSE current is obtained at finite temperature. This is because then the limit $N_t \to \infty$ means the transition to continuum limit, and the appropriate tuning of the lattice spacing $a$ allows to treat the final answer as the axial current at finite temperature $T = 1/(N_t a)$. However, for the interacting system the situation may be different, and at finite temperatures the corrections to the CSE current may appear \cite{Puhr:2016kzp} - the effect, which we do not discuss here.

We conclude, that in the physical regularizations with Wilson and overlap fermions the conventional CSE emerges. At the same time we suppose, that the model with the modified overlap fermions \cite{overlap} with exact chiral symmetry Eq. (\ref{exactchiral}) is unphysical. Although the zeros of the Green function do not contribute to the ordinary perturbation expansion on the same grounds as the physical excitations, they contribute the topological quantities responsible for the CSE in the same way as the fermion doublers.  In this respect the modified overlap fermions with Eq. (\ref{exactchiral}) are similar to the naive lattice fermions with $16$ doublers, and they do not possess the CSE.

\section*{Acknowledgements}
MAZ kindly acknowledges numerous discussions with G.E.Volovik. K.V.Z. is greatful to A.Yu.Kotov for discussions. Both authors are greatful to M.N.Chernodub for useful discussions, and to LMPT, the University of Tours, where this work was initiated, for kind hospitality. The present work was supported by Russian Science Foundation Grant No 16-12-10059.

\section*{Appendix A. Axial current for the modified overlap fermions}

Here we calculate the chiral current for the version of overlap fermions with exact chiral symmetry $\{{\cal G},\gamma^5\}=0$. Let us use the following expression for the axial current
\begin{eqnarray}
j^{5k}&=&-\frac{i}{2}\sum_{n=1}^{N_t}\int \frac{d^3p}{(2\pi)^3}Tr(\gamma^5{\cal G}(\omega_{n},\textbf{p})\partial_{p_{i}}{\cal G}^{-1}(\omega_{n},\textbf{p})\nonumber \\&&\partial_{p_{j}}{\cal G}(\omega_{n},\textbf{p})\partial_{p_{k}}{\cal G}^{-1}(\omega_{n},\textbf{p}))F_{ij}
\end{eqnarray}
We substitute Eq. (\ref{eqs}) to this expression:
\begin{eqnarray}
j^{5k}&=&-\frac{i}{2}\sum_{n=1}^{N_t}\int\frac{d^3p}{(2\pi)^3}Tr(\gamma^{5}\frac{\gamma^{\mu}k_{\mu}}{k^2 f(k^2)}\partial^{i}(\gamma^{\nu}k_{\nu}f(k^2))\nonumber \\&&\partial^{j}(\frac{\gamma^{\lambda}k_{\lambda}}{k^2 f(k^2)})(\partial^{k}\gamma^{\rho}k_{\rho}f(k^2)))F_{ij}
\end{eqnarray}
 It may be written as
\begin{equation}
j^{5k}=-2i\sum_{n=1}^{N_t}\epsilon^{\rho \mu \nu \lambda}\int_{M}{\frac{d^3p}{(2\pi)^3} \frac{f^2(k^2) k_{\mu}\partial^{i} k_{\nu}\partial^{j}k_{\lambda}\partial^{k}k_{\rho}}{f^2(k^2)k^4}}F_{ij}
\end{equation}
 We introduce the notation $g^{\mu}=\frac{k^{\mu}}{\sqrt{k^2}}$.
And the expression for the axial current is given by
\begin{equation}
j^{5k}=-\frac{2i}{3! (2\pi)^3}\sum_{n=1}^{Nt}\epsilon^{\mu\nu\lambda\rho}\int_{M}{g_{\mu}dg_{\nu}\wedge dg_{\lambda} \wedge dg_{\rho}}\epsilon^{ijk}F_{ij}
\end{equation}

To calculate the topological invariant
\begin{equation}{\cal N}_{3}(\omega_n)=\frac{1}{12 \pi^2}\epsilon^{\mu\nu\lambda\rho}\int_{M}{g_{\mu}dg_{\nu}\wedge dg_{\lambda} \wedge dg_{\rho}}\end{equation}
we will use the method of \cite{Z2016_1}, and the following parametrization:
\begin{equation}
g_{4}={\rm sin}(\alpha), g_{i}=k_{i}{\rm cos}(\alpha)
\end{equation}
where $i=1,2,3$ and $\sum_{i}k^{2}_{i}=1$, while $\alpha  \in [\frac{-\pi}{2},\frac{\pi}{2}]$. Thus
\begin{equation}dg_{4}={\rm cos}(\alpha)d\alpha
,dg_{i}=dk_i {\rm cos}(\alpha)-k_{i}{\rm sin}(\alpha)d\alpha \end{equation}
 And
$$
{\cal N}_{3}=\frac{3}{12 \pi^2}\epsilon^{ijk}\int_{M}{{\rm cos}^2(\alpha)k_i d\alpha \wedge d k_j \wedge d k_k}$$
$$
=\frac{3}{12 \pi^2}\epsilon^{ijk}\int_{M}{k_i(\frac{1-{\rm cos}(2\alpha)}{2})d\alpha \wedge d k_{j} \wedge d k_k}=$$

$$
=\frac{3}{12 \pi^2}\epsilon^{ijk}\int_{M}{k_i d(\frac{\alpha}{2}+\frac{1}{4}{\rm sin}(\alpha)) \wedge d k_{j} \wedge d k_{k}}$$
\begin{equation}
=-\sum_{l}\frac{3}{12 \pi^2}\epsilon^{ijk}\int_{\partial \Omega}{k_{i}(\frac{\alpha}{2}+\frac{1}{4}{\rm sin}(\alpha)) d k_{j} \wedge d k_{k}}
\end{equation}
In this  expression $\partial \Omega$ is the  small vicinity of point $y_l$ of momentum space  where vector $k_l$ is undefined. The absence of the {\rm sin}gularities of $g_{k}$ implies that $\alpha \to \pm \frac{\pi}{2}$ at such points. \par
Thus we see that the expression under the integral is the total derivative. We can rewrite it  in  the form

\begin{equation}
{\cal N}_{3}=-\frac{1}{2}\sum_{l}{\rm sign}(g_{4}(y_{l})){\rm Res}(y_{l})
\end{equation}
Where  we have used the notation \cite{Z2016_2}
\begin{equation} {\rm Res}(y_l)=\frac{1}{8\pi}\epsilon^{ijk}\int_{\partial \Omega}{g_i dg_j \wedge dg_k}\end{equation}
It is worth mentioning, that this symbol obeys $\sum_{l}{\rm Res}(y_l)=0$. At each $n$ the value of ${\rm sign} \, g_4$ is constant. Therefore, ${\cal N}_3(\omega_n) = 0$ for any $n$.

\end{document}